\documentclass[fleqn,10pt]{wlscirep}
\usepackage[utf8]{inputenc}
\usepackage[T1]{fontenc}
\usepackage{siunitx}

\title{Stain Consistency Learning: Handling Stain Variation for Automatic Digital Pathology Segmentation}

\author[1,*]{Michael Yeung}
\author[2]{Todd Watts}
\author[3]{Sean YW Tan}
\author[4]{Pedro F. Ferreira}
\author[4]{Andrew D. Scott}
\author[4]{Sonia Nielles-Vallespin}
\author[5,6,7]{Guang Yang}
\affil[1]{Imperial College London, Department of Computing, London, SW7 2AZ, UK}
\affil[2]{The Babraham Institute, Signalling Programme, Cambridge, CB22 3AT, UK}
\affil[3]{University of Cambridge, Department of Clinical Neurosciences, Cambridge, CB2 2PY, UK}
\affil[4]{National Heart \& Lung Institute, Imperial College London, London, SW3 6LY, UK}
\affil[5]{Bioengineering Department and Imperial-X, Imperial College London, London W12 7SL, UK}
\affil[6]{Cardiovascular Research Centre, Royal Brompton Hospital, London SW3 6NP, UK}
\affil[7]{School of Biomedical Engineering \& Imaging Sciences, King's College London, London WC2R 2LS, UK}

\affil[*]{michael.yeung21@imperial.ac.uk}

\keywords{Data Augmentation, Deep Learning, Digital Pathology, Instance Segmentation, Stain Variation}

\begin{abstract}
Stain variation is a unique challenge associated with automated analysis of digital pathology. Numerous methods have been developed to improve the robustness of machine learning methods to stain variation, but comparative studies have demonstrated limited benefits to performance. Moreover, methods to handle stain variation were largely developed for H\&E stained data, with evaluation generally limited to classification tasks. Here we propose Stain Consistency Learning, a novel framework combining stain-specific augmentation with a stain consistency loss function to learn stain colour invariant features. We perform the first, extensive comparison of methods to handle stain variation for segmentation tasks, comparing ten methods on Masson's trichrome and H\&E stained cell and nuclei datasets, respectively. We observed that stain normalisation methods resulted in equivalent or worse performance, while stain augmentation or stain adversarial methods demonstrated improved performance, with the best performance consistently achieved by our proposed approach. The code is available at: \href{https://github.com/mlyg/stain_consistency_learning}{https://github.com/mlyg/stain\_consistency\_learning}.
\end{abstract}
\begin{document}

\flushbottom
\maketitle

\thispagestyle{empty}

\section*{Introduction}

With the increasing demand for tissue biopsies and the number pathologists remaining in short supply \cite{robboy2013pathologist}, there is an growing need for automatic methods to process histopathology data. In particular, there has been recent interest in artificial intelligence (AI)-based computational pathology \cite{cui2021artificial}, which provides automatic, accurate and efficient interpretation of digital pathology \cite{baxi2022digital,hanna2022integrating,niazi2019digital,cui2021artificial}, with performance comparable to expert pathologists \cite{alsubaie2018bottom, lu2020prognostic}.

Despite the significant progress made with AI-based computational pathology,
there remains several legal, ethical and technical challenges that limit its translation into clinical practice \cite{berbis2023computational}. Among the technical challenges, stain variation represents one of the most important barriers hindering the generalisation of machine learning models to external data. This is because machine learning models may overfit to stain appearances observed on training data, and fail to generalise to images from other centres or scanners which display different staining characteristics \cite{ciompi2017importance}. 

\begin{figure*}[ht!]
    \includegraphics[scale=0.75]{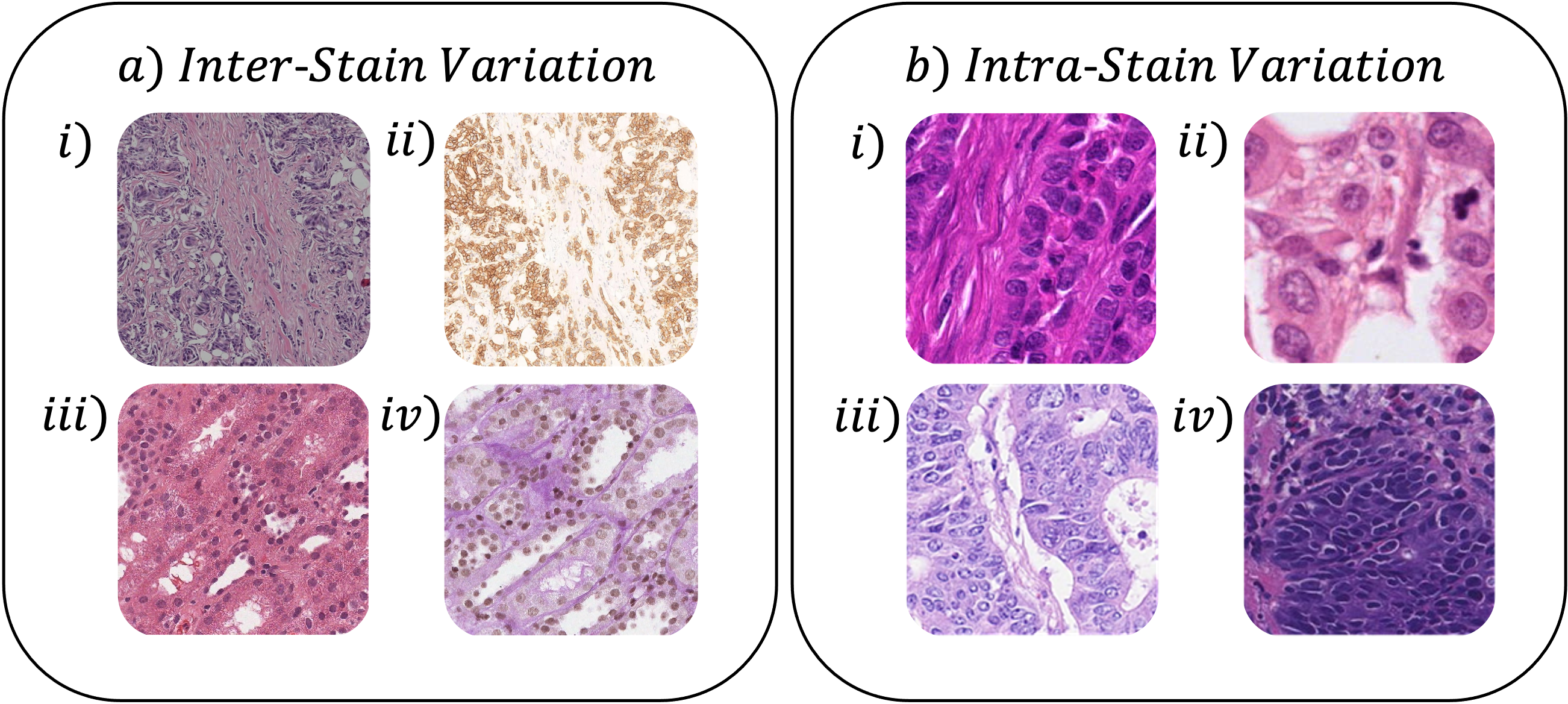}
    \centering
    \caption{Examples of stain variation in digital pathology. a) Inter-stain variation of consecutive i) H\&E and ii) HER2 receptor antibody stained data, and iii) H\&E and iv) Periodic acid–Schiff (PAS) stained data. b) Intra-stain variation of H\&E stained data. Inter-stain variation images were selected from Breast Cancer Immunohistochemical (BCI) dataset (a i-ii) and Center for Applied Medical Research (CIMA) histology dataset (a iii-iv), and intra-stain variation images were selected from the MIDOG dataset (b i-ii) and Lizard dataset (b iii-iv).}
    \label{fig:figure_1}
\end{figure*}

Stains are chemical dyes applied to histology slides to enhance tissue contrast and highlight certain structures. The most widely used stain is haematoxylin and eosin (H\&E), which stains nuclei dark purple and the extracellular matrix and cytoplasm pink. For specific situations, other stains may be preferred, such as Masson's trichrome stain for connective tissue, or immunohistochemical stains for proteins. Depending on the stain used, the same tissue section can differ considerably in appearance (Figure ~\ref{fig:figure_1}).

Alongside inter-stain variation, there is also intra-stain variation, where tissue appearances may differ despite applying the same stain. Stain variation arises during the complex process of acquiring digital pathology, where differences in tissue preparation protocol, stain manufacturer and scanner type may contribute towards both inter-stain and intra-stain variation \cite{roy2018study}.

\begin{figure*}[ht!]
    \includegraphics[scale=0.6]{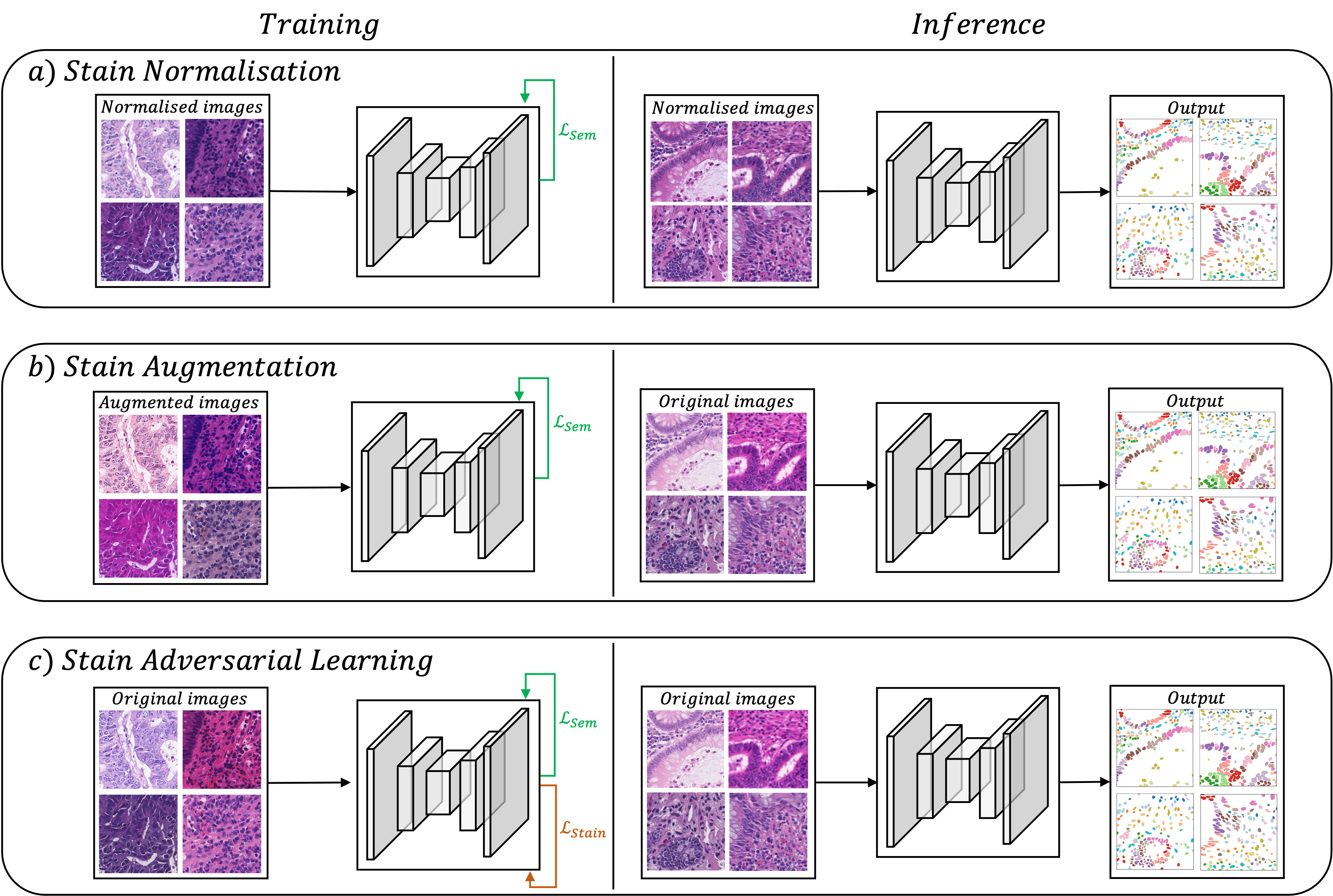}
    \centering
    \caption{Overview of methods to handle stain variation on digital pathology. a) Stain normalisation methods standardise the staining prior to training and inference, removing the need for machine learning models to learn stain variation. b) Stain augmentation methods apply stain-specific perturbations to the training data, increasing the diversity of staining observed by the model during training. c) Stain adversarial learning involves joint optimisation of a specialised loss function ($\mathcal{L}_\text{Stain}$) to learn stain-invariant features in addition to the downstream task ($\mathcal{L}_\text{Sem}$).}
    \label{fig:figure_2}
\end{figure*}

In recent years, there has been considerable progress in developing methods to improve the robustness of machine learning models to stain variation \cite{roy2018study,tellez2019quantifying}. These methods can be broadly categorised into stain normalisation, stain augmentation and stain adversarial learning approaches (Figure ~\ref{fig:figure_2}). 

Stain normalisation handles stain variation by standardising the staining prior to model training and inference, reducing the total variation in staining seen by the model. Broadly, stain normalisation approaches can be divided into statistical, stain separation and deep learning methods.

Statistical methods involve matching the image characteristics of a reference image to a set of images and include histogram matching (HM) \cite{neumann2005color,coltuc2006exact,gonzalez2009digital}, Fourier Domain Adaptation (FDA) \cite{yang2020fda, yang2022source, aubreville2022mitosis}, and Reinhard's method \cite{reinhard2001color}. These methods can be applied to a wide variety of imaging data but are prone to generating artefacts or require careful hyperparameter tuning. 

Stain separation methods were developed specifically for histology data, and involve matching the stain characteristics of a reference image to a set of images. Ruifrok and Johnston proposed the first stain separation method, known as the colour deconvolution (CD) algorithm \cite{ruifrok2001quantification}, which decomposes an image into stain colour and stain concentration components. However, CD is not scalable because it requires obtaining the stain colour matrix empirically. Later approaches focused on automatically extracting the stain colour matrix through matrix decomposition, including the use of singular value decomposition (SVD) \cite{macenko2009method}, non-negative matrix factorisation (NMF) \cite{rabinovich2003unsupervised, vahadane2016structure} and integrated optimisation \cite{zheng2019adaptive}.

In recent years, deep learning-based methods using generative adversarial networks (GANs) have become a popular choice for performing stain normalisation, reframing the problem as an unsupervised, style-transfer task \cite{goodfellow2020generative, bentaieb2017adversarial, shaban2019staingan, kang2021stainnet}. There has been progress in reducing the inference time \cite{kang2021stainnet}, as well as adapting GANs to handle multiple domains \cite{choi2018stargan,nazki2023multipathgan}. However, it is currently unclear whether GANs are suitable methods for clinical use because there are uncertainties as to whether the original image content could be altered during the transformation \cite{cohen2018distribution}.

While stain normalisation must be applied during inference, stain augmentation and stain adversarial approaches are only necessary during training. Stain augmentations are colour augmentation methods that were developed for histology images \cite{tellez2018whole, wagner2021structure, shen2022randstainna}. Stain Jitter generates diversity by applying noise directly to the H\&E colour channels extracted using CD \cite{tellez2018whole}, with two hyperparameters controlling the degree of pertubation. RandStainNA is a more recent approach that overcomes the need for hyperparameter tuning by using the training data to determine the degree of stain perturbation \cite{shen2022randstainna}. The use of GANs to generate stain diversity has also been explored \cite{wagner2021structure,vasiljevic2021towards,scalbert2022test}, although existing GAN-based augmentation approaches are computationally expensive and require domain labels for training \cite{li2023laplacian}.

Stain adversarial methods encourage optimisation for learning stain-invariant features using specialised loss functions. Domain-adversarial neural network (DANN) training involves the use of a task-specific loss, such as classification or segmentation, together with a domain classification loss \cite{otalora2019staining}. To avoid the need for domain labels, Marini et al. replaced the domain classification loss with a regression loss and trained models to maximise the regression loss when predicting H\&E colour matrices \cite{marini2021h}. However, stain adversarial methods rely on network modifications for generating auxiliary outputs, limiting their flexibility for integration into existing pipelines.

Despite numerous methods developed to handle stain variation, many of the methods have not been widely adopted due to problems associated with computational complexity, need for domain labels or hyperparameter tuning, and concern over image content alterations. Moreover, the majority of methods to handle stain variation were developed and evaluated on classification tasks using H\&E-stained data and it is unclear whether these methods generalise to different stains or tasks. 

To address these issues, we propose a novel framework, named stain consistency learning (SCL), to handle variation of different stains for segmentation tasks. In this paper, we propose the following contributions:

\begin{enumerate}[label=(\alph*)]
  \item We propose stain consistency learning, a novel method that incorporates stain-specific augmentation with a specialised loss function to encourage the learning of stain colour invariant features.
  \item We perform the first large-scale comparison of methods to handle stain variation for segmentation tasks, evaluating ten methods on datasets of Masson's trichrome stained cells and H\&E stained nuclei, and show that stain consistency learning outperforms all other methods across datasets.
  \item We demonstrate the benefit of leveraging larger, unlabelled datasets to improve the performance of stain augmentation methods, alleviating the dependency on stain variation within the training data.
\end{enumerate}

The paper is structured as follows. Section 1 introduces our proposed stain consistency learning method, and describes the datasets and implementation details. Section 2 presents the experimental results. Finally, Section 3 discusses the results and provides conclusive remarks and future directions. 

\section{Material and methods}

\subsection{Stain Consistency Learning}

SCL incorporates a new stain augmentation method to increase stain diversity observed by the model, and a complementary loss function to encourage the learning of stain colour invariant features.

\subsubsection{Stain Consistency Augmentation}

Both Tellez et al. and RandStainNA use a single, empirically determined stain colour matrix to compute the stain concentration matrix. However, a single colour matrix does not account for intra-stain variation, which may affect the colour deconvolution process if the stain appearance of the image differs significantly from the reference colour matrix.

\begin{figure*}[ht!]
    \includegraphics[scale=0.5]{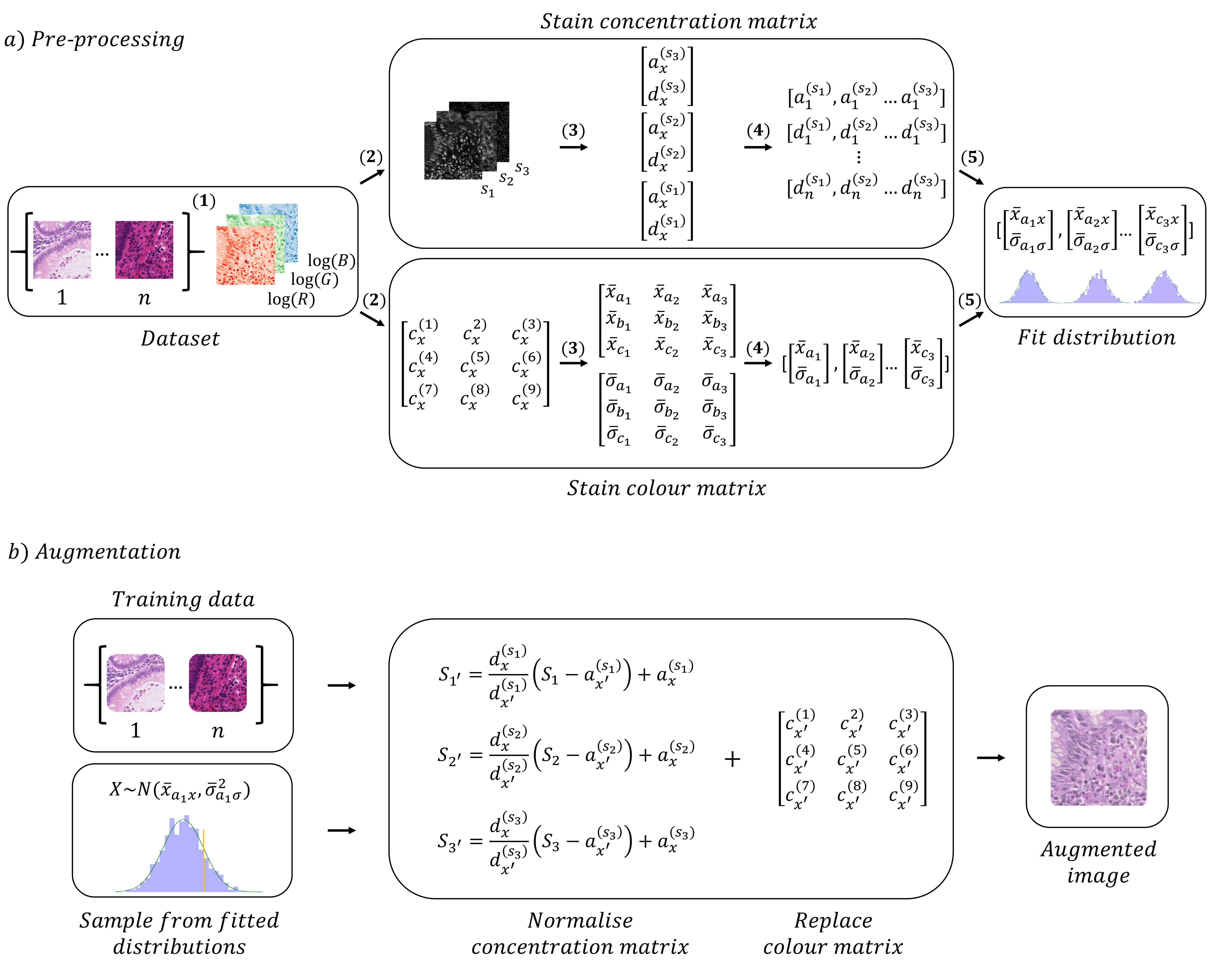}
    \centering
    \caption{Overview of Stain Consistency Augmentation. a) Pre-processing. 1) Each image is first converted into optical density space. 2) Images are decomposed into a stain colour and stain concentration matrix using Macenko's method. 3) The mean and standard deviation statistics are extracted per channel for the stain concentration matrix, and per element for the stain colour matrix. 4) Mean and standard deviation statistics are aggregated across the dataset. 5) A normal distribution is fit for each statistic. b) Augmentation. During training, the selected image is decomposed into stain colour and stain concentration matrices. The stain colour matrix is replaced with one sampled from the fitted distributions, while the per-channel mean and standard deviation statistics of the stain concentration matrix are normalised using the sampled stain concentration values, generating an image with augmented stain colour and concentration properties.}
    \label{fig:figure_5}
\end{figure*}

Therefore, we instead propose modelling each image as a unique stain appearance, and extract the stain colour matrix from each image independently. An overview of our approach, stain consistency augmentation (SCA), is shown in Figure ~\ref{fig:figure_5}.

To augment an image (I), the image is first converted to optical density (OD) space:

\begin{equation}
\text I_{OD}= -\text{log}_{10} (\text I).
\end{equation}

In OD space, the image can be decomposed into a stain colour matrix (\textbf{C}) and stain concentration matrix (\textbf{S}):

\begin{equation}
\text I_{OD}= \mathbf{C}\mathbf{S}.
\end{equation}

Macenko's method was used to obtain the stain colour matrix \cite{macenko2009method}. The stain concentration matrix can then be obtained by inverting the stain colour matrix:

\begin{equation}
\mathbf{S} = \mathbf{C}^{-1} \text I_{OD}.
\end{equation}

For each image, we compute the stain colour matrix $\mathbf{C}_i \in \mathbb{R}^{3 \times 3}$, as well as the mean $\mathbf{A}_i=\left[a_i^{(s_{1})}, a_i^{(s_{2})}, a_i^{(s_{3})}\right] \in \mathbb{R}^{3}$ and standard deviations $\mathbf{D}_i=\left[d_i^{(s_{1})}, d_i^{(s_{2})}, d_i^{(s_{3})}\right] \in \mathbb{R}^{3}$ for each stain $s \in \{s_{1}, s_{2}, s_{3} \}$ in the stain concentration matrix.

Following this, we adopt RandStainNA's approach by fitting a distribution over the dataset for the extracted features \cite{shen2022randstainna}. Specifically, we fit three multivariate Gaussian distributions $F_{C}$, $F_{A}$, and $F_{D}$, for the stain colour matrix, as well as the mean and standard deviation of the stain concentration matrix, respectively:

\begin{equation}
\begin{aligned}
F_{C} \sim \mathcal{N}(\boldsymbol{M}_{C}, \boldsymbol{\Sigma}_{C}), \\
F_{A} \sim \mathcal{N}(\boldsymbol{M}_{A}, \boldsymbol{\Sigma}_{A}), \\
F_{D} \sim \mathcal{N}(\boldsymbol{M}_{D}, \boldsymbol{\Sigma}_{D}).
\end{aligned}
\end{equation}

To augment a given image $\text I_{x}$, we first extract the stain colour matrix $\mathbf{C}_x \in \mathbb{R}^{3 \times 3}$, as well as the per-channel mean $\mathbf{A}_x=\left[a_x^{(s_{1})}, a_x^{(s_{2})}, a_x^{(s_{3})}\right]$ and standard deviation $\mathbf{D}_x=\left[d_x^{(s_{1})}, d_x^{(s_{2})}, d_x^{(s_{3})}\right] \in \mathbb{R}^{3}$ of the stain concentration matrix. Next, we sample from the fitted distributions to obtain a stain colour matrix $\mathbf{C}_{x'} \in \mathbb{R}^{3 \times 3}$, and sample mean $\mathbf{A}_{x'}=\left[a_{x'}^{(s_{1})}, a_{x'}^{(s_{2})}, a_{x'}^{(s_{3})}\right]$ and standard deviations $\mathbf{D}_{x'}=\left[d_{x'}^{(s_{1})}, d_{x'}^{(s_{2})}, d_{x'}^{(s_{3})}\right] \in \mathbb{R}^{3}$ to normalise the stain concentration matrix.
We replace the stain colour matrix $\mathbf{C}_x$ with $\mathbf{C}_{x'}$, and transform the stain concentration matrix $\mathbf{S}$:

\begin{equation}
\begin{aligned}
S_{1'} & =\frac{d_x^{(s_{1})}}{d_{x'}^{(s_{1})}}\left(s_{1}-a_{x'}^{(s_{1})}\right)+a_x^{(s_{1})} \\
S_{2'} & =\frac{d_x^{(s_{2})}}{d_{x'}^{(s_{2})}}\left(s_{2}-a_{x'}^{(s_{2})}\right)+a_x^{(s_{2})} \\
S_{3'} & =\frac{d_x^{(s_{3})}}{d_{x'}^{(s_{3})}}\left(s_{3}-a_{x'}^{(s_{3})}\right)+a_x^{(s_{3})}.
\end{aligned}
\end{equation}

Finally, the modified stain colour and stain concentration matrices are combined and converted to RGB space, producing the augmented image, $\text I_{x'}$:

\begin{equation}
\text I_{x'}= e^{-\mathbf{C}_{x'}\mathbf{S}_{x'}}.
\end{equation}

\subsubsection{Stain Consistency Loss}

\begin{figure*}[ht!]
    \includegraphics[scale=0.45]{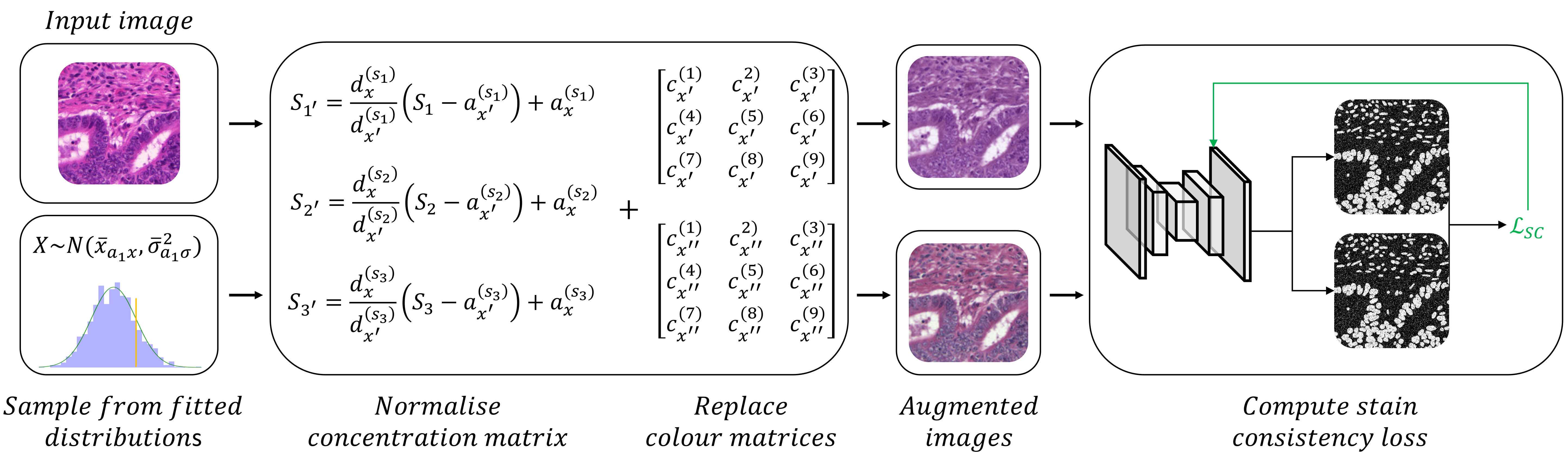}
    \centering
    \caption{Overview of Stain Consistency Learning. For each training image, we sample a single stain concentration matrix and two stain colour matrices from our fitted distributions. These are used to generate two augmented variants, by normalising the original image with the same stain concentration matrix and replacing the stain colour matrix with different stain colour matrices. Both augmented variants are then input into the model separately, and the stain consistency loss is computed based on the two segmentation outputs.}
    \label{fig:figure_6}
\end{figure*}

Stain augmentation methods encourage the learning of stain-invariant features by requiring model predictions to remain accurate in the presence of stain variation. These methods are not readily compatible with stain adversarial learning approaches because augmentations may fail to preserve domain-specific variation. To bridge the gap between stain augmentation and stain adversarial learning approaches, we instead propose SCL (Figure ~\ref{fig:figure_6}).

Using SCA, it is possible to generate stain colour and concentration variation. SCL involves the optimisation of machine learning models to learn features that are invariant to stain colour while remaining sensitive to stain concentration. This is useful because the variation in stain colour represents the intra-stain differences that should not affect model predictions, while variation in stain concentration represents important structural differences that models should discriminate. 

To learn stain colour invariance while preserving sensitivity to stain concentration, we apply SCA to generate two augmented images with the same stain concentration transformation but different stain colour matrices. By passing both images as inputs, we define the stain consistency loss ($\mathcal{L}_\text{SC}$) as the mean absolute error between the two segmentation outputs, $p$ and $p'$, excluding background pixels and following transformation using a sigmoid activation function:

\begin{equation}
\mathcal{L}_\text{SC}(p, p')= \frac{1}{m} \sum_{i \in M} |p_{i} - p'_{i}|,
\label{eq:sc}
\end{equation}

where m iterates over the pixels corresponding to the ground truth objects.

\subsection{Dataset description}

\subsubsection{Lizard Dataset}

The Lizard dataset is the largest open-source dataset with instance segmentation labels in digital pathology, comprised of 495,179 labeled nuclei from six sources \cite{graham2021lizard}. We use the training image patches of the Lizard dataset prepared by the CoNIC challenge \cite{graham2021conic} (Table S1). 

The Lizard dataset consists of H\&E-stained whole slide images of colonic tissue at 20$\times$ objective magnification. Nuclei instance segmentation masks were generated through a semi-automatic approach, using HoVer-Net to provide annotations that were subsequently manually refined for challenging cases \cite{graham2019hover}. The CoNIC challenge provides 4,981 256$\times$256 non-overlapping patches, from which we remove all patches without nuclei, leaving 4,841 patches for training and evaluation. 

\subsubsection{Cardiomyocyte Dataset}

Our in-house dataset consists of four porcine hearts, arrested in a diastolic-like state, that were subsequently formalin fixed and paraffin embedded \cite{nielles2017assessment}. Transmural sections were obtained at 5$\mu$m thickness and stained with Masson's trichrome. We obtained high-resolution microscopy data at 20$\times$ magnification with 456nm/pixel resolution (C9600-12, Hamamatsu, Hamamatsu City, Japan). We randomly extracted 200 224 $\times$ 224 patches per heart, with two sets of cell instance segmentation labels separately provided by two individuals experienced with histology data.

\subsection{Implementation details}

We used the VersaTile framework for instance segmentation \cite{versatile}, with the default training settings. The data augmentation settings for VersaTile are shown in Table S2.

For cell segmentation, we perform four-fold cross validation, involving successive training on images from one subject, and evaluation on images from the other three subjects. For the Lizard dataset, we use image patches from DigestPath for training, and image patches from CoNSeP, CRAG, GlaS and PanNuke for evaluation. We selected DigestPath for training because it is the only dataset within the Lizard dataset that does not have an overlapping dataset source with the other datasets. For each method, we train three models, initialised with different random seeds, and report the average performance.

All experiments were conducted on a single NVIDIA RTX-6000 GPU and programmed in PyTorch. For evaluating current methods, we used the official implementations where available (FDA, ACD, Stain Jitter, RandStainNA), and the popular StainTools library for Reinhard's method, Macenko's method and SPCN.

To test for statistical significance, we used the Wilcoxon signed-rank test. A statistically significant difference was defined as $p<\num{0.05}$.

\section{Results}

\subsection{Qualitative comparison of stain augmentation methods}

To provide a qualitative comparison of different stain augmentation methods, we provide a range of examples of augmented H\&E images from the Lizard dataset and Masson's trichrome images from our in-house dataset in Figure ~\ref{fig:figure_3}. The appearance of stain normalised images are available in the supplementary materials (Figure S1). 

\begin{figure*}[ht!]
    \includegraphics[scale=0.5]{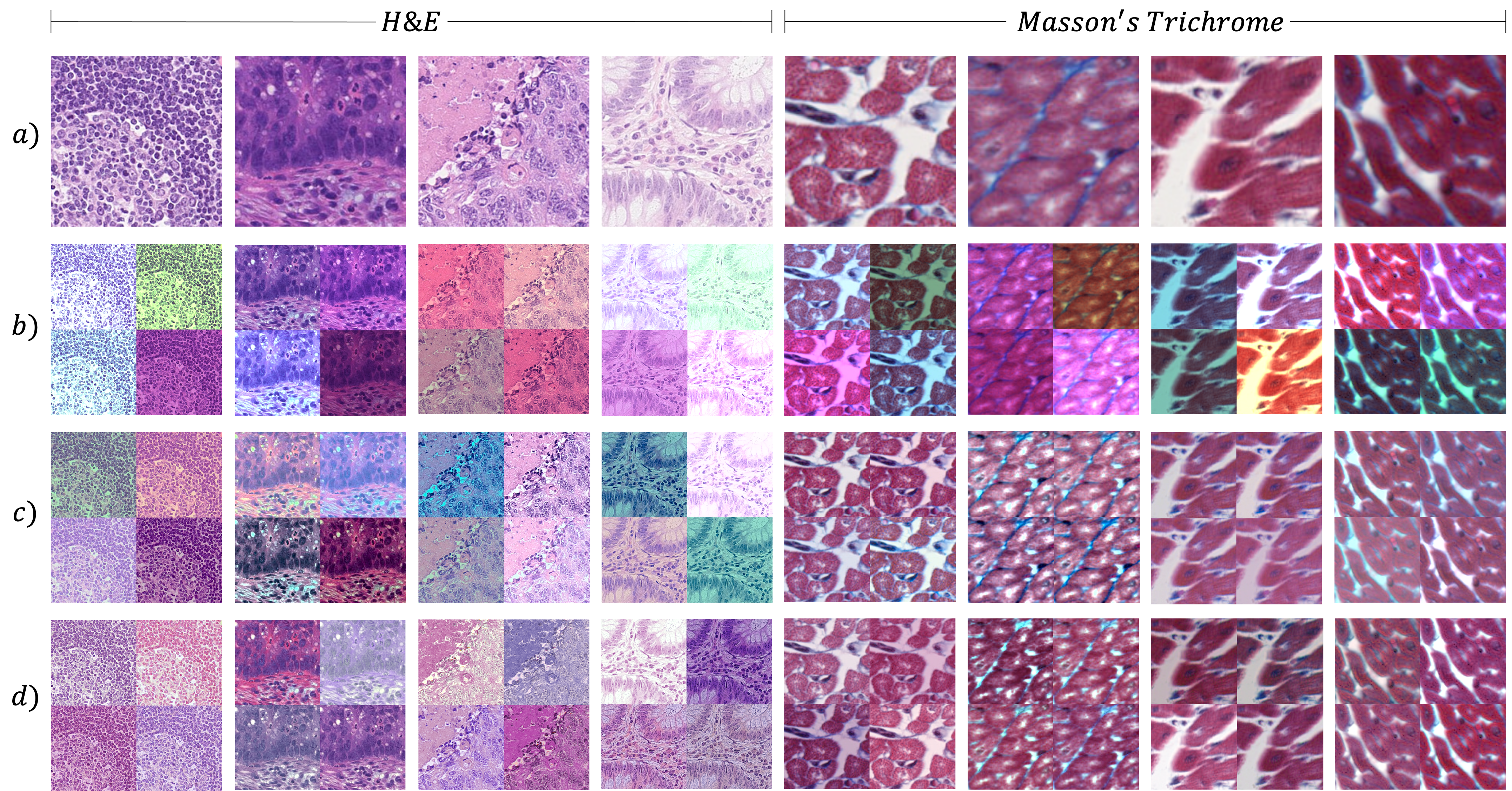}
    \centering
    \caption{Example images after applying different stain augmentation methods. a) Original image. b) Stain Jitter \cite{tellez2018whole}. c) RandStainNA ($L\alpha\beta$ space) \cite{shen2022randstainna}. d) SCA (proposed). The images were selected from the Lizard dataset and our in-house dataset to represent a range of stain appearances. Each augmented image demonstrates four consecutive applications of the stain augmentation method on the original image. RandStainNA and SCA parameters were fit on the Lizard dataset. For Stain Jitter, we set $\alpha$ = 0.25 and $\beta$ = 0.05 \cite{tellez2018whole}.}
    \label{fig:figure_3}
\end{figure*}

As shown in Figure ~\ref{fig:figure_3}, stain augmentation methods are able to generate a diverse range of stain appearances. While Stain Jitter appears to produce the most diverse range of colour variation, it also generates the largest number of unrealistic augmented images. This is particularly evident with poorly stained images, where the addition of noise could result in stain colour values shifting to outside of the realistic range. In contrast, both RandStainNA and SCA do not display systematic errors and maintain generally realistic stain variation across the sample of images. However, applying RandStainNA occasionally results in the introduction of colour artefacts, which is not apparent with SCA, and overall SCA generates the most realistic stain variation.

\subsection{Quantitative comparison on instance segmentation}

We evaluate the effects of using various stain normalisation, stain augmentation and stain adversarial learning methods on cell and nuclei segmentation performance.

The results for cell segmentation are shown in Table ~\ref{fig:table_4}.

\begin{table*}[ht]
\centering
\caption{Performance on cell segmentation using different methods to handle stain variation. The best performance is denoted in bold. The bootstrapped 95\% confidence intervals are shown in brackets. The overall performance is computed as the mean performance using annotator 1 and 2 as ground truth labels.}
\scalebox{0.9}{
\begin{tabular}{lcccccc}
\cline{2-7}
                       & \multicolumn{2}{c}{Annotator 1}                              & \multicolumn{2}{c}{Annotator 2}                     & \multicolumn{2}{c}{Overall}              \\ \hline
Method        & $F1_{50}$ ($\uparrow$)              & $PQ_{50}$ ($\uparrow$)                                  & $F1_{50}$ ($\uparrow$)                        & $PQ_{50}$ ($\uparrow$)             & $F1_{50}$ ($\uparrow$)             & $PQ_{50}$ ($\uparrow$)             \\ \hline
Baseline               & 0.735 ($\pm$0.005)          & 0.576 ($\pm$0.005)                              & 0.736 ($\pm$0.005)                     & 0.586 ($\pm$0.005)          & 0.736 ($\pm$0.005)          & 0.581 ($\pm$0.005)          \\
HM \cite{coltuc2006exact}                     & 0.725 ($\pm$0.005)          & 0.564 ($\pm$0.005)                              & 0.723 ($\pm$0.005)                     & 0.572 ($\pm$0.005)          & 0.724 ($\pm$0.005)          & 0.568 ($\pm$0.005)          \\
FDA \cite{yang2020fda}                    & 0.729 ($\pm$0.005)          & 0.568 ($\pm$0.005)                              & 0.729 ($\pm$0.005)                     & 0.579 ($\pm$0.005)          & 0.729 ($\pm$0.005)          & 0.574 ($\pm$0.005)          \\
Reinhard et al. \cite{reinhard2001color}        & 0.734 ($\pm$0.005)          & 0.573 ($\pm$0.005)                              & 0.736 ($\pm$0.005)                     & 0.584 ($\pm$0.005)          & 0.735 ($\pm$0.005)          & 0.579 ($\pm$0.005)          \\
Macenko et al. \cite{macenko2009method}        & 0.723 ($\pm$0.005)          & 0.566 ($\pm$0.005)                              & 0.715 ($\pm$0.005)                     & 0.561 ($\pm$0.005)          & 0.719 ($\pm$0.005)          & 0.564 ($\pm$0.005)          \\
SPCN \cite{vahadane2016structure}       & 0.722 ($\pm$0.005)          & 0.560 ($\pm$0.005)                              & 0.722 ($\pm$0.005)                     & 0.567 ($\pm$0.005)          & 0.722 ($\pm$0.005)          & 0.564 ($\pm$0.005)          \\
ACD \cite{zheng2019adaptive}                   & 0.732 ($\pm$0.005)          & 0.573 ($\pm$0.005)                              & 0.729 ($\pm$0.005)                     & 0.579 ($\pm$0.005)          & 0.731 ($\pm$0.005)          & 0.576 ($\pm$0.005)          \\
Stain Jitter \cite{tellez2018whole}          & 0.739 ($\pm$0.005)          & 0.580 ($\pm$0.005)                              & 0.737 ($\pm$0.005)                     & 0.587 ($\pm$0.005)          & 0.738 ($\pm$0.005)          & 0.584 ($\pm$0.005)          \\
RandStainNA \cite{shen2022randstainna}           & 0.739 ($\pm$0.005)          & 0.580 ($\pm$0.005)                              & 0.741 ($\pm$0.005)                     & 0.591 ($\pm$0.005)          & 0.740 ($\pm$0.005)          & 0.586 ($\pm$0.005)          \\ \hline
SCA (proposed)  & 0.745 ($\pm$0.005)          & 0.586 ($\pm$0.005)                              & \multicolumn{1}{l}{0.745 ($\pm$0.005)} & 0.595 ($\pm$0.005)          & 0.745 ($\pm$0.005)          & 0.591 ($\pm$0.005)          \\
SCL (proposed) & \textbf{0.754 ($\pm$0.005)} & \multicolumn{1}{l}{\textbf{0.594 ($\pm$0.005)}} & \textbf{0.757 ($\pm$0.005)}            & \textbf{0.606 ($\pm$0.005)} & \textbf{0.756 ($\pm$0.005)} & \textbf{0.600 ($\pm$0.005)} \\ \hline
\end{tabular}}
\label{fig:table_4}
\end{table*}

Overall, the best performing method was SCL, with an F1 score of 0.756 ($\pm$0.005) and PQ score of 0.600 ($\pm$0.005). This was followed by SCA, with an overall F1 score of 0.745 ($\pm$0.005) and PQ score of 0.591 ($\pm$0.005). All stain augmentation methods performed significantly better than the baseline for both F1 scores (Stain Jitter: $p=\num{9.2e-9}$, RandStainNA: $p=\num{0.017}$, SCA: $p=\num{9.2e-9}$, SCL: $p<\num{1e-10}$) and PQ scores (Stain Jitter: $p=\num{3.1e-10}$, RandStainNA: $p=\num{5.0e-3}$, SCA: $p=\num{3.1e-10}$, SCL: $p<\num{1e-10}$). In contrast, stain normalisation approaches were associated with equivalent F1 (Reinhard: $p=\num{0.18}$, SPCN: $p=\num{0.12}$) and PQ (Reinhard: $p=\num{0.12}$) scores, or worse F1 (HM: $p=\num{2.9e-3}$, FDA: $p=\num{0.014}$, Macenko: $p=\num{0.0014}$, ACD: $p=\num{2.4e-7}$) and PQ (HM: $p=\num{1.8e-3}$, FDA: $p=\num{0.0077}$, Macenko: $p=\num{5.3e-5}$, SPCN: $p=\num{0.0023}$, ACD: $p=\num{1.8e-9}$) scores.

The results for nuclei segmentation are shown in Table ~\ref{fig:table_5}.

\begin{table*}[ht]
\centering
\caption{Performance on nuclei segmentation using different methods to handle stain variation. The best performance is denoted in bold. The bootstrapped 95\% confidence intervals are shown in brackets. The overall performance is the mean performance across the four datasets.}
\scalebox{0.6}{
\begin{tabular}{lcccccccccc}
\cline{2-11}
\textbf{}              & \multicolumn{2}{c}{CoNSeP}              & \multicolumn{2}{c}{CRAG}                 & \multicolumn{2}{c}{GlaS}                 & \multicolumn{2}{c}{PanNuke}              & \multicolumn{2}{c}{Overall}              \\ \hline
Method        & $F1_{50}$ ($\uparrow$)            & $PQ_{50}$ ($\uparrow$)            & $F1_{50}$ ($\uparrow$)            & $PQ_{50}$ ($\uparrow$)            & $F1_{50}$ ($\uparrow$)            & $PQ_{50}$ ($\uparrow$)            & $F1_{50}$ ($\uparrow$)            & $PQ_{50}$ ($\uparrow$)            & $F1_{50}$ ($\uparrow$)            & $PQ_{50}$ ($\uparrow$)            \\ \hline
Baseline               & 0.671 ($\pm$0.010)          & 0.505 ($\pm$0.008)          & 0.614 ($\pm$0.003)          & 0.471 ($\pm$0.002)          & 0.691 ($\pm$0.003)          & 0.531 ($\pm$0.003)          & 0.711 ($\pm$0.009)          & 0.539 ($\pm$0.007)          & 0.672 ($\pm$0.006)          & 0.512 ($\pm$0.005)          \\
HM \cite{coltuc2006exact}                    & 0.645 ($\pm$0.012)          & 0.479 ($\pm$0.010)          & 0.567 ($\pm$0.003)          & 0.430 ($\pm$0.003)          & 0.668 ($\pm$0.004)          & 0.507 ($\pm$0.003)          & 0.676 ($\pm$0.011)          & 0.511 ($\pm$0.009)          & 0.639 ($\pm$0.008)          & 0.482 ($\pm$0.006)          \\
FDA \cite{yang2020fda}                    & 0.597 ($\pm$0.015)          & 0.438 ($\pm$0.011)          & 0.506 ($\pm$0.004)          & 0.382 ($\pm$0.003)          & 0.657 ($\pm$0.004)          & 0.496 ($\pm$0.003)          & 0.624 ($\pm$0.013)          & 0.465 ($\pm$0.01)           & 0.596 ($\pm$0.009)          & 0.445 ($\pm$0.007)          \\
Reinhard et al. \cite{reinhard2001color}       & 0.646 ($\pm$0.011)          & 0.482 ($\pm$0.009)          & 0.603 ($\pm$0.003)          & 0.460 ($\pm$0.002)          & 0.666 ($\pm$0.004)          & 0.510 ($\pm$0.003)          & 0.692 ($\pm$0.009)          & 0.522 ($\pm$0.007)          & 0.652 ($\pm$0.007)          & 0.494 ($\pm$0.005)          \\
Macenko et al. \cite{macenko2009method}        & 0.631 ($\pm$0.013)          & 0.472 ($\pm$0.010)          & 0.518 ($\pm$0.004)          & 0.391 ($\pm$0.003)          & 0.665 ($\pm$0.003)          & 0.503 ($\pm$0.003)          & 0.674 ($\pm$0.011)          & 0.507 ($\pm$0.009)          & 0.622 ($\pm$0.008)          & 0.468 ($\pm$0.006)          \\
SPCN \cite{vahadane2016structure} & 0.613 ($\pm$0.013)          & 0.458 ($\pm$0.010)          & 0.522 ($\pm$0.004)          & 0.394 ($\pm$0.003)          & 0.667 ($\pm$0.004)          & 0.504 ($\pm$0.003)          & 0.640 ($\pm$0.013)          & 0.477 ($\pm$0.010)          & 0.611 ($\pm$0.009)          & 0.458 ($\pm$0.007)          \\
ACD \cite{zheng2019adaptive}                   & 0.644 ($\pm$0.013)          & 0.484 ($\pm$0.010)          & 0.527 ($\pm$0.004)          & 0.403 ($\pm$0.003)          & 0.702 ($\pm$0.003)          & 0.538 ($\pm$0.003)          & 0.691 ($\pm$0.011)          & 0.523 ($\pm$0.009)          & 0.641 ($\pm$0.008)          & 0.487 ($\pm$0.006)          \\
StainGAN \cite{shaban2019staingan}              & 0.500 ($\pm$0.016)          & 0.362 ($\pm$0.012)          & 0.552 ($\pm$0.003)          & 0.414 ($\pm$0.002)          & 0.540 ($\pm$0.004)          & 0.388 ($\pm$0.003)          & 0.568 ($\pm$0.013)          & 0.420 ($\pm$0.010)          & 0.540 ($\pm$0.009)          & 0.396 ($\pm$0.007)          \\
Stain Jitter \cite{tellez2018whole}          & 0.692 ($\pm$0.009)          & 0.521 ($\pm$0.008)          & 0.677 ($\pm$0.002)          & 0.518 ($\pm$0.002)          & 0.704 ($\pm$0.003)          & 0.530 ($\pm$0.003)          & 0.720 ($\pm$0.007)          & 0.547 ($\pm$0.006)          & 0.698 ($\pm$0.005)          & 0.529 ($\pm$0.005)          \\
RandStainNA \cite{shen2022randstainna}           & 0.682 ($\pm$0.009)          & 0.513 ($\pm$0.008)          & 0.613 ($\pm$0.003)          & 0.469 ($\pm$0.002)          & 0.677 ($\pm$0.003)          & 0.520 ($\pm$0.003)          & 0.715 ($\pm$0.008)          & 0.543 ($\pm$0.007)          & 0.672 ($\pm$0.006)          & 0.511 ($\pm$0.005)          \\
H\&E Adversarial \cite{marini2021h}       & 0.680 ($\pm$0.010)          & 0.509 ($\pm$0.008)          & 0.639 ($\pm$0.003)          & 0.489 ($\pm$0.002)          & 0.683 ($\pm$0.003)          & 0.524 ($\pm$0.003)          & 0.719 ($\pm$0.009)          & 0.546 ($\pm$0.007)          & 0.680 ($\pm$0.006)          & 0.517 ($\pm$0.005)          \\ \hline
SCA (proposed)  & 0.695 ($\pm$0.010)          & 0.524 ($\pm$0.008)          & \textbf{0.685 ($\pm$0.002)} & \textbf{0.526 ($\pm$0.002)} & 0.719 ($\pm$0.003)          & 0.545 ($\pm$0.003)          & 0.723 ($\pm$0.008)          & 0.549 ($\pm$0.007)          & 0.706 ($\pm$0.006)          & 0.536 ($\pm$0.005)          \\
SCL (proposed) & \textbf{0.696 ($\pm$0.010)} & \textbf{0.525 ($\pm$0.008)} & 0.683 ($\pm$0.002)          & 0.524 ($\pm$0.002)          & \textbf{0.726 ($\pm$0.003)} & \textbf{0.553 ($\pm$0.002)} & \textbf{0.730 ($\pm$0.007)} & \textbf{0.555 ($\pm$0.007)} & \textbf{0.709 ($\pm$0.006)} & \textbf{0.539 ($\pm$0.005)} \\ \hline
\end{tabular}}
\label{fig:table_5}
\end{table*}

Similar to cell segmentation, SCL was the best performing method for nuclei segmentation, with an overall F1 score of 0.709 ($\pm$0.006) and PQ score of 0.539 ($\pm$0.005). This was followed by SCA with an overall F1 score of 0.706 ($\pm$0.006) and PQ score of 0.536 ($\pm$0.006). Both stain augmentation and stain adversarial methods performed significantly better than the baseline for both F1 and PQ scores (all $p<\num{1e-10}$). In contrast, all stain normalisation approaches were associated with significantly lower overall F1 and PQ scores (all $p<\num{1e-10}$). 

\subsection{Effect of pre-processing datasets on stain augmentation}

Both RandStainNA and SCA involve learning a distribution over the images in the training data to generate stain diversity. Rather than using the training data, it is possible to learn the stain distribution using a larger unlabelled dataset, which would allow models to learn stain variation not present in the training data. To evaluate the effect of using different datasets for pre-processing on segmentation performance, we pre-process RandStainNA and SCA on the MIDOG dataset and the whole Lizard dataset \cite{aubreville2022mitosis}. The results are shown in Table ~\ref{fig:table_6}.

\begin{table*}[ht]
\centering
\caption{Performance on nuclei segmentation using different pre-processing datasets for stain augmentation methods. DigestPath is the training data used in the other experiments and is shown for reference. The best performances are denoted in bold. The bootstrapped 95\% confidence intervals are shown in brackets. The overall performance is computed as the mean performance across the four datasets.}
\scalebox{0.55}{
\begin{tabular}{llcccccccccc}
\cline{2-12}
               &            & \multicolumn{2}{c}{CoNSeP}                        & \multicolumn{2}{c}{CRAG}                          & \multicolumn{2}{c}{GlaS}                          & \multicolumn{2}{c}{PanNuke}                       & \multicolumn{2}{c}{Overall}                       \\ \hline
Method         & Dataset    & $F1_{50}$ ($\uparrow$)                      & $PQ_{50}$ ($\uparrow$)                     & $F1_{50}$ ($\uparrow$)                      & $PQ_{50}$ ($\uparrow$)                     & $F1_{50}$ ($\uparrow$)                      & $PQ_{50}$                      & $F1_{50}$ ($\uparrow$)                      & $PQ_{50}$ ($\uparrow$)                     & $F1_{50}$ ($\uparrow$)                      & $PQ_{50}$ ($\uparrow$)                     \\ \hline
RandStainNA    & DigestPath & 0.682 ($\pm$0.009)          & 0.513 ($\pm$0.008)          & 0.613 ($\pm$0.003)          & 0.469 ($\pm$0.002)          & 0.677 ($\pm$0.003)          & 0.520 ($\pm$0.003)          & 0.715 ($\pm$0.008)          & 0.543 ($\pm$0.007)          & 0.672 ($\pm$0.006)          & 0.511 ($\pm$0.005)          \\
SCA (proposed) & DigestPath & 0.695 ($\pm$0.010)          & 0.524 ($\pm$0.008)          & \textbf{0.685 ($\pm$0.002)} & \textbf{0.526 ($\pm$0.002)} & 0.719 ($\pm$0.003)          & 0.545 ($\pm$0.003)          & 0.723 ($\pm$0.008)          & 0.549 ($\pm$0.007)          & 0.706 ($\pm$0.006)          & 0.536 ($\pm$0.005)          \\
SCL (proposed) & DigestPath & \textbf{0.696 ($\pm$0.010)} & \textbf{0.525 ($\pm$0.008)} & 0.683 ($\pm$0.002)          & 0.524 ($\pm$0.002)          & \textbf{0.726 ($\pm$0.003)} & \textbf{0.553 ($\pm$0.002)} & \textbf{0.730 ($\pm$0.007)} & \textbf{0.555 ($\pm$0.007)} & \textbf{0.709 ($\pm$0.006)} & \textbf{0.539 ($\pm$0.005)} \\ \hline
RandStainNA    & MIDOG      & 0.676 ($\pm$0.010)          & 0.508 ($\pm$0.008)          & 0.625 ($\pm$0.003)          & 0.479 ($\pm$0.002)          & 0.685 ($\pm$0.003)          & 0.527 ($\pm$0.003)          & 0.708 ($\pm$0.009)          & 0.538 ($\pm$0.008)          & 0.674 ($\pm$0.006)          & 0.513 ($\pm$0.005)          \\
SCA (proposed) & MIDOG      & 0.693 ($\pm$0.011)          & 0.523 ($\pm$0.009)          & \textbf{0.700 ($\pm$0.002)} & \textbf{0.537 ($\pm$0.002)} & 0.715 ($\pm$0.003)          & 0.540 ($\pm$0.003)          & \textbf{0.726 ($\pm$0.008)} & \textbf{0.551 ($\pm$0.007)} & 0.709 ($\pm$0.006)          & 0.538 ($\pm$0.005)          \\
SCL (proposed) & MIDOG      & \textbf{0.696 ($\pm$0.011)} & \textbf{0.524 ($\pm$0.009)} & 0.695 ($\pm$0.002)          & 0.533 ($\pm$0.002)          & \textbf{0.723 ($\pm$0.003)} & \textbf{0.552 ($\pm$0.003)} & 0.725 ($\pm$0.007)          & \textbf{0.551 ($\pm$0.007)} & \textbf{0.710 ($\pm$0.006)} & \textbf{0.540 ($\pm$0.005)} \\ \hline
RandStainNA    & Lizard     & 0.680 ($\pm$0.010)          & 0.512 ($\pm$0.008)          & 0.608 ($\pm$0.003)          & 0.466 ($\pm$0.002)          & 0.677 ($\pm$0.003)          & 0.521 ($\pm$0.003)          & \textbf{0.719 ($\pm$0.008)}          & \textbf{0.545 ($\pm$0.007)} & 0.671 ($\pm$0.006)          & 0.511 ($\pm$0.005)          \\
SCA (proposed) & Lizard     & \textbf{0.697 ($\pm$0.011)} & \textbf{0.525 ($\pm$0.009)} & \textbf{0.696 ($\pm$0.002)} & \textbf{0.535 ($\pm$0.002)} & 0.726 ($\pm$0.003)          & 0.554 ($\pm$0.003)          & 0.715 ($\pm$0.010) & 0.544 ($\pm$0.008)          & \textbf{0.709 ($\pm$0.007)} & \textbf{0.540 ($\pm$0.006)} \\
SCL (proposed) & Lizard     & 0.692 ($\pm$0.010)          & 0.519 ($\pm$0.009)          & 0.692 ($\pm$0.002)          & 0.531 ($\pm$0.002)          & \textbf{0.727 ($\pm$0.009)} & \textbf{0.555 ($\pm$0.002)} & 0.715 ($\pm$0.011) & 0.543 ($\pm$0.007)          & 0.707 ($\pm$0.008)          & 0.537 ($\pm$0.005)          \\ \hline
\end{tabular}}
\label{fig:table_6}
\end{table*}

Overall, we observed better segmentation performance using images from the MIDOG or Lizard dataset for stain augmentation pre-processing, compared to using images from the training data. SCA performance significantly improved from an F1 score of 0.706 ($\pm$0.006) and PQ score of 0.536 ($\pm$0.005), to an F1 score of 0.709 ($\pm$0.007) and PQ score of 0.540 ($\pm$0.006), when using the Lizard dataset for pre-processing ($p<\num{1e-10}$). Specifically, the CRAG subset was associated with the greatest performance improvement, where the F1 score improved from 0.685 ($\pm$0.002) to 0.696 ($\pm$0.002), and PQ score improved from 0.526 ($\pm$0.002) to 0.535 ($\pm$0.002). Less consistent performance improvements were observed using SCL and RandStainNA. The best overall performance was associated with SCL pre-processed on MIDOG, with an F1 score improvement to 0.710 ($\pm$0.006) ($p=\num{0.072}$) and PQ score improvement to 0.540 ($\pm$0.005) ($p=\num{1.9e-4}$).


\section{Discussion}

Stain variation is a unique and significant challenge for developing automated methods to process digital pathology. In this study, we performed the first, extensive comparison of methods to handle stain variation in the context of instance segmentation. Moreover, we proposed a novel approach, named SCL, that integrates stain augmentation with a specialised loss function to learn stain colour invariant features. 

Our results are in-line with previous results from a multi-centre study which systematically evaluated the effects of stain normalisation and stain augmentation on four classification tasks \cite{tellez2019quantifying}. Specifically, we observe improved performance using stain augmentation and adversarial methods (Tables ~\ref{fig:table_4} and ~\ref{fig:table_5}), but equivalent or worse performance using stain normalisation approaches (Tables ~\ref{fig:table_4} and ~\ref{fig:table_5}), despite the inclusion of newer methods such as StainGAN (Table ~\ref{fig:table_5}). While the majority of stain normalisation methods were developed and evaluated on classification tasks, instance segmentation is significantly more challenging, requiring both pixel-level accuracy and understanding of object boundaries. Instance segmentation is therefore more sensitive to small structural changes to the image that may be introduced by applying normalisation approaches (Figure S1), which suggests that stain normalisation may not be appropriate for segmentation tasks. 

We proposed a novel stain augmentation method named SCA that extends upon the methods of Stain Jitter and RandStainNA. Rather than using a single, empirically determined stain colour matrix, we used Macenko's method to individually extract stain colour matrices for each image. We leveraged the diversity in stain colour matrices to generate realistic stain-specific variation (Figure ~\ref{fig:figure_3}), and observed performance benefits over other methods to handle stain variation across Masson's trichrome-stained cardiomyocytes (Table ~\ref{fig:table_4}) and H\&E-stained nuclei segmentation (Table ~\ref{fig:table_5}). Instead of using the training data for preprocessing, we explored learning the stain distribution from external data and observed performance benefits (Table ~\ref{fig:table_6}). This provides a convenient mechanism to overcome limited stain diversity in the training data by learning the stain distribution on a larger, more diverse dataset, without the need for labels.

To integrate stain augmentation and stain adversarial learning for handling stain variation, we developed stain consistency learning, a framework that encourages the model to learn stain colour invariant features while preserving sensitivity to stain concentration differences. This method demonstrated the best performance overall across both datasets (Tables ~\ref{fig:table_4} and ~\ref{fig:table_5}).  

Alongside performance, SCL has several favourable properties compared to the other methods developed to handle stain variation. Firstly, neither our proposed augmentation approach nor loss function require hyperparameter tuning. This is in contrast to other stain augmentation approaches such as Stain Jitter, which has two hyperparameters \cite{tellez2018whole}, and stain normalisation approaches such ACD, which has five hyperparameters \cite{zheng2019adaptive}. Secondly, SCL is only applied during training and no modifications are required during inference. This not only avoids increasing inference time, but also avoids structural changes to the images. Thirdly, SCL does not require modifications to the neural network architecture and can therefore be more easily integrated into existing frameworks, in contrast to stain adversarial approaches which require an additional auxiliary output.

There are also several limitations to SCL. Firstly, to use the stain consistency loss, two augmented variants of an image are passed to the network with each iteration, which increases the overall training time. However, SCA is computationally fast (Table S3), and the training time could be further reduced by pre-computing the stain colour matrices, which could also allow otherwise computationally expensive but more reliable methods such as SPCN to be used for computing stain colour matrices. Secondly, we use Macenko's method for extracting stain colour matrices, which may occasionally generate structural artefacts. While robustness to structural artefacts is crucial for stain normalisation methods, small structural artefacts as part of stain augmentation may provide benefit through a regularisation effect. Lastly, our method of learning a distribution to generate realistic stain augmentation relies on pre-processing using a dataset with diverse stain appearances. This is because the specific distribution fit is highly dependent on the stain composition of the dataset.

For future work, there are several areas where SCL can be extended. Firstly, to facilitate the learning of a distribution that represents diverse stain appearances, either a balanced dataset could be curated, or methods such as sampling or clustering could be used to avoid over-representation of common stain appearances. Secondly, it would be useful to evaluate our framework on other image recognition tasks, to determine whether our method can also provide performance benefits to other tasks. Finally, we demonstrated that SCL improves robustness to intra-stain variation, but it would be useful to extend this framework to also handle inter-stain variation.

\bibliography{references}

\section{Acknowledgements}
Guang Yang was supported in part by the ERC IMI (101005122), the H2020 (952172), the MRC (MC/PC/21013), the Royal Society (IEC \textbackslash NSFC \textbackslash 211235), the NVIDIA Academic Hardware Grant Program, the SABER project supported by Boehringer Ingelheim Ltd, and the UKRI Future Leaders Fellowship (MR/V023799/1).


\section{Author contributions statement}

M.Y and G.Y. conceived the experiments, P.F., A.S., S.N., T.W. and S.T. prepared the dataset, M.Y. conducted the experiments, M.Y. analysed the results. All authors reviewed the manuscript.

\section{Data availability statement}

The nuclei segmentation dataset was obtained from the Colon Nuclei Identification and Counting (CoNIC) Challenge and is publicly available at \href{https://conic-challenge.grand-challenge.org/}{https://conic-challenge.grand-challenge.org/}. The cell segmentation dataset is not currently publicly available, but may be available from the corresponding author upon reasonable request and with permission from the National Heart \& Lung Institute, Imperial College London.

\section{Additional information}

\textbf{Competing interests}

The authors declare no competing interests.


\end{document}


\section{Stain normalisation examples}
\label{A}

\begin{figure*}[ht!]
    \includegraphics[scale=0.45]{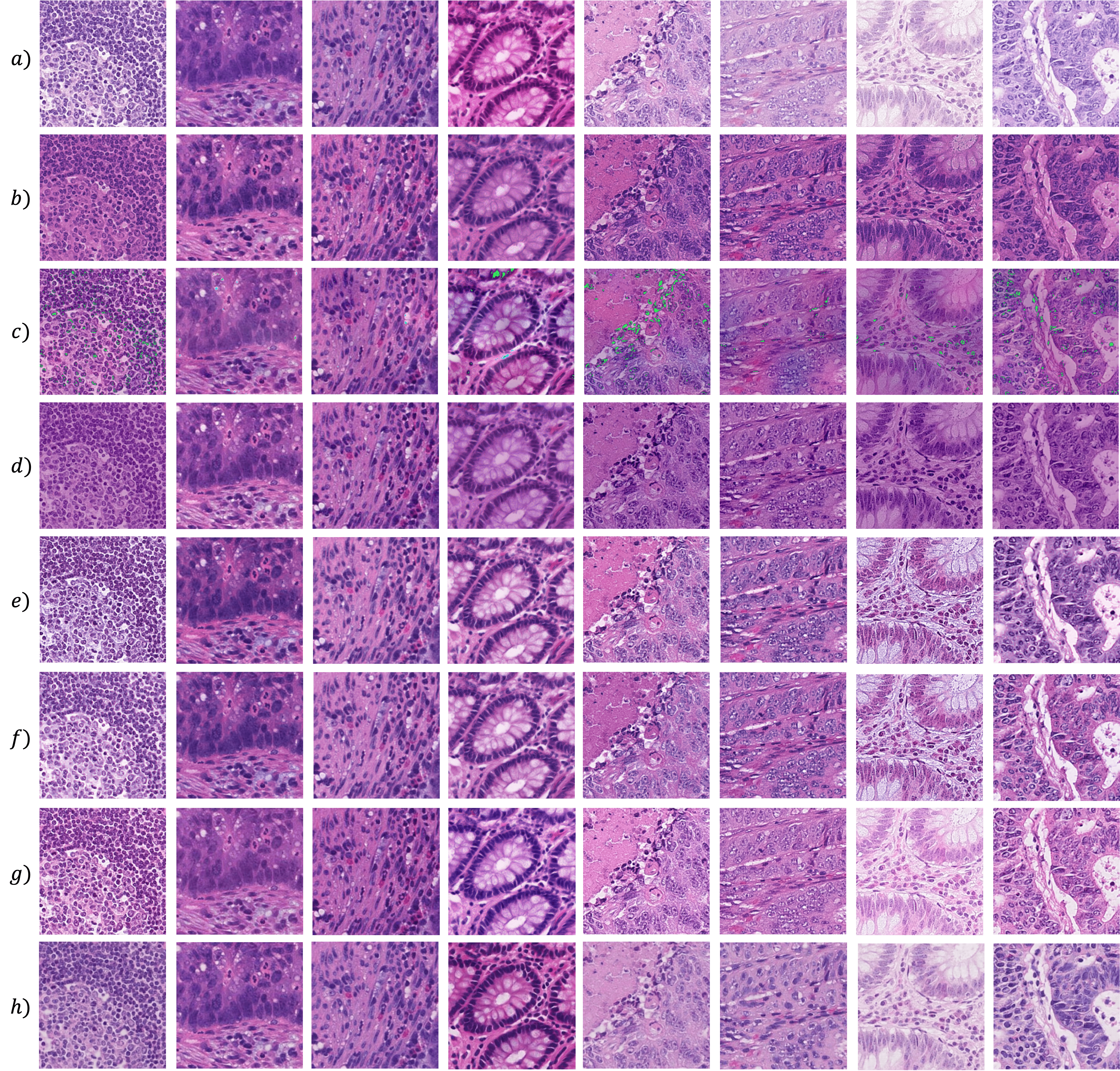}
    \centering
    \caption{Example images after applying different stain normalisation methods. a) Original image. b) HM \cite{coltuc2006exact}. c) FDA \cite{yang2020fda}. d) Reinhard et al. \cite{reinhard2001color}. e) Macenko et al. \cite{macenko2009method}. f) SPCN \cite{vahadane2016structure}. g) ACD \cite{zheng2019adaptive}. h) StainGAN \cite{shaban2019staingan}. The images were carefully selected from the Lizard dataset to represent a range of stain appearances. Where applicable, normalised images used the same reference image.}
    \label{fig:figure_4}
\end{figure*}

\newpage

\begin{table*}[ht]
\centering
\caption{Lizard dataset sources. All datasets except for DigestPath contain data from University Hospitals Coventry and Warwickshire (UHCW). The sixth dataset source in Lizard, The Cancer Genome Atlas Program (TCGA), was not included in the Colon Nuclei Identification and Counting Challenge (CoNIC) challenge training data.}
\scalebox{1.}{
\begin{tabular}{lccc}
\textbf{Dataset} & \textbf{Source}                                                             & \textbf{\#Images} & \multicolumn{1}{l}{\textbf{\#Annotations}} \\ \hline
DigestPath       & Multiple centres in China                                                   & 1,788              & 225,859                                     \\ 
CoNSeP           & UHCW                                                                        & 64                & 6,595                                       \\ 
CRAG             & UHCW                                                                        & 2,181              & 210,957                                     \\ 
GlaS             & UHCW                                                                        & 698               & 112,292                                     \\ 
PanNuke          & \begin{tabular}[c]{@{}c@{}}Multiple centres in USA and \\ UHCW\end{tabular} & 110               & 14,158                                      \\ \hline
\end{tabular}}
\label{fig:table_1}
\end{table*}

\newpage

\begin{table*}[ht]
\centering
\caption{Data augmentation settings. All augmentations used the default settings in Albumentations version 1.3.1. One of CLAHE, RandomGamma and RandomBrightnessContrast was used per augmentation.}
\scalebox{1.}{
\begin{tabular}{ll}
\hline
\textbf{Augmentation}    & \multicolumn{1}{c}{\textbf{Setting}}               \\ \hline
RandomResizedCrop        & scale=(0.08, 1), ratio=(0.75, 1.3), p=0.25         \\
HorizontalFlip           & p=0.5                                              \\
VerticalFlip             & p=0.5                                              \\
RandomRotation90         & p=0.5                                              \\
CLAHE                    & p=0.25                                             \\
RandomGamma              & gamma\_limit=(80,120), p=0.25                      \\
RandomBrightnessContrast & brightness\_limit=0.2, contrast\_limit=0.2, p=0.25 \\
Blur                     & blur\_limit=7, p=0.25                              \\ \hline
\end{tabular}}
\label{fig:table_3}
\end{table*}

\newpage

\begin{table*}[ht]
\centering
\caption{Processing time using different stain normalisation and augmentation methods. Processing time is reported using a CPU to process 1000 images of size $256 \times 256$.}
\scalebox{1}{
\begin{tabular}{lc}
\hline
Method                         & Processing Time (s) ($\downarrow$) \\ \hline
HM \cite{coltuc2006exact}                      & 154                 \\
FDA \cite{yang2020fda}                        & 35                  \\
Reinhard et al. \cite{reinhard2001color}                       & 4                   \\
Macenko et al. \cite{macenko2009method}                        & 234                 \\
SPCN \cite{vahadane2016structure}                    & 2183                \\
ACD \cite{zheng2019adaptive}                           & 312                 \\
Stain Jitter \cite{tellez2018whole}                  & 5                   \\
RandStainNA \cite{shen2022randstainna}                   & 4                   \\
SCA (proposed) & 15                  \\ \hline
\end{tabular}}
\label{fig:table_7}
\end{table*}

\newpage

\bibliography{references}